\documentclass[a4paper,11pt]{article}
\usepackage{pos}

\usepackage[FIGBOTCAP,TABBOTCAP,tight]{subfigure}
\usepackage{tikz}
\usepackage{graphicx}
\usepackage{grffile}
\usepackage{overpic}
\usepackage{subfigure}
\usepackage[utf8]{inputenc}
\usepackage{amsmath}
\graphicspath{{./fig/}}
\newcommand{\expval}[1]{\langle {#1} \rangle}
\title{Error reduction using machine learning on Ising worm simulation}

\author*[a]{Jangho Kim}
\author[b]{Wolfgang Unger}

\affiliation[a]{Institute for Advanced Simulation (IAS-4), Forschungszentrum J\"ulich, \\ Wilhelm-Johnen-Stra\ss e, 52428 J\"ulich, Germany}
\affiliation[b]{Fakult\"at f\"ur Physik, Universit\"at Bielefeld, \\
Universit\"atstra\ss e 25, D33619 Bielefeld, Germany}

\emailAdd{j.kim@fz-juelich.de}
\emailAdd{wunger@physik.uni-bielefeld.de}

\abstract{We develop a method to improve on the statistical errors for higher moments
using machine learning techniques.
We present here results for the dual representation of the Ising model
with an external field, derived via the high temperature expansion and simulated by the worm algorithm.
We compare two ways of measuring the same set of observables, without and with machine learning: moments of the magnetization and the susceptibility can be improved by using the decision tree method to train the correlations between the higher moments and the second moment obtained from an integrated 2-point function. Those results are compared in small volumes to analytic predictions.
}

\FullConference{%
The 39th International Symposium on Lattice Field Theory,\\
8th-13th August, 2022,\\
Rheinische Friedrich-Wilhelms-Universität Bonn, Bonn, Germany
}

\begin{document}
\maketitle

\section{Ising dual representation}
The dual representation of the Ising model is derived by introducing bond variables and integrating out the spin degrees of freedom \cite{Prokof:2001,Wolff2008,Gabriel:2002}: 
\begin{align}
	Z_{\rm Ising}&=\sum_{\{s\}} e^{-\beta H(s)},\qquad H=-J \sum_{\langle i,j \rangle}s_i s_j+h\sum_i s_i\nonumber\\
&=(2 \cosh(\beta h))^{V} \cosh(\beta J)^{E}  \sum_{\stackrel{\{n_b,m_i\}}{\partial \{n_b\} = \{m_i\}}}\tanh(\beta J)^{\sum_b n_b}\tanh(\beta h)^{\sum_i m_i},
\end{align}
where the first line is the standard spin representation of the Ising model and the second line is its dual representation derived by the high temperature character expansion in $\beta=1/T$, and $h$ is the external magnetic field. 
The dual variables of that representation are the monomers $m_i\in \{0,1\}$ defined on the lattice sites $i$ and the dimers $n_b \in \{0,1\}$ defined on the bonds $b$, which are nearest neighbor pairs. Here, $V$ is the volume, $E$ is the number of bonds. Since spin summation only gives non-trivial contributions if after the expansion the spin at every site is raised to an even power, we obtain the constraint that the dimers form intersecting loops that are either closed or form strings that connect two monomers: $\partial \{n_b\} = \{m_i\}$.
The dual representation is well suited to be simulated via Monte Carlo, in particular using the worm algorithm \cite{Prokof:2001}. Then we can measure the number of monomers $M=\sum_{i} m_i$ and its higher moments on each configuration obtained after a worm update, but also so-called improved estimators, such as the 2-point correlation functions during worm evolution \cite{Wolff2008}. 
It turns out that the averaged worm length is simply the connected susceptibility obtained via the integrated 2-point function
\begin{align}
\langle G_2 \rangle&=\frac{1}{V^2}\sum_{x,y}  \langle G(x,y)\rangle  =\langle \sigma^2 \rangle 
\end{align}
 with $G(x,y)=G(x-y)$ due to translation symmetry.
\begin{figure}[b]
	\subfigure[The second moment $\expval{\sigma^2}$]{
        \includegraphics[width=0.49\textwidth]{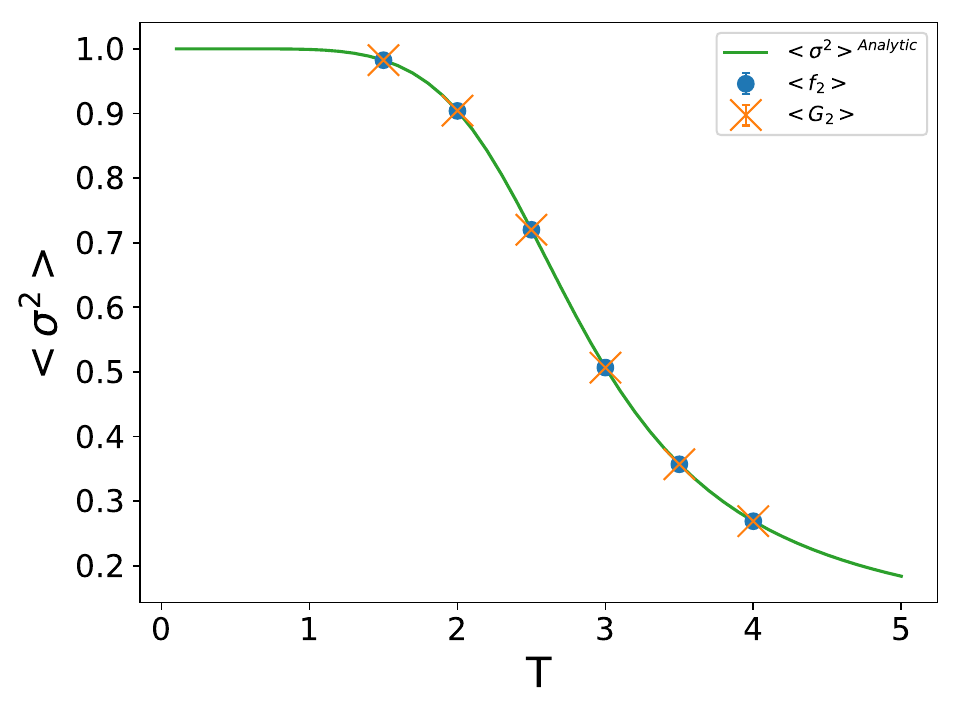}
			}
	\subfigure[Deviation from the exact solution]{\label{fig:s2_sub}
				\includegraphics[width=0.49\textwidth]{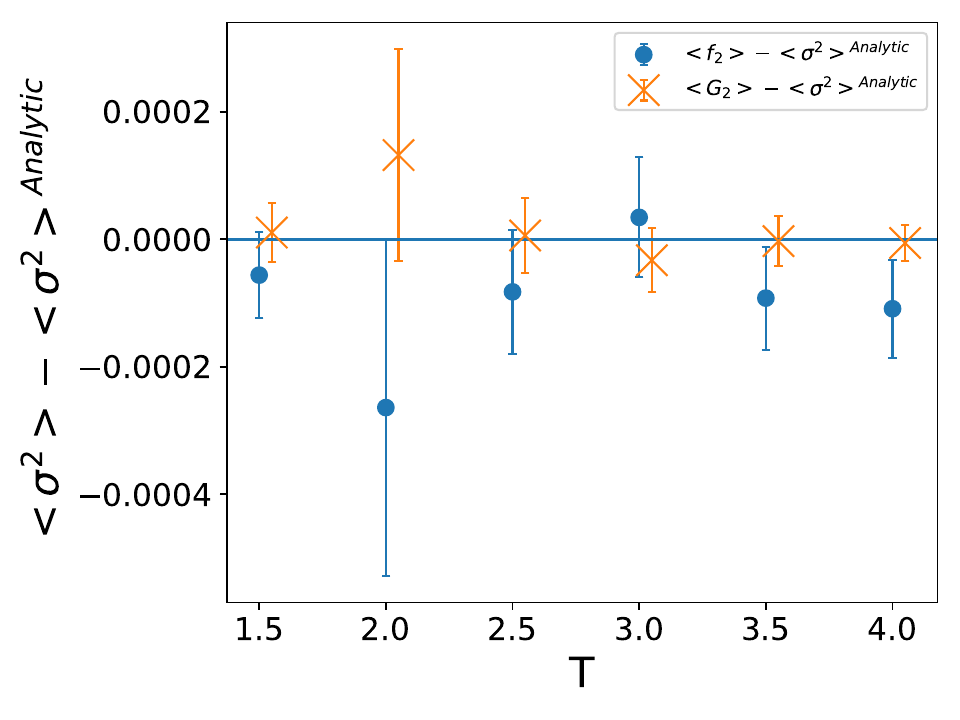}
	}
	\caption{\label{fig:s2}
	Left: Comparison of $\expval{f_2}$ and $\expval{G_2}$ with analytic solution on a $4 \times 4$ lattice and the external field $h=0.2$. Right: Subtracting the exact solution from $\expval{f_2}$ and $\expval{G_2}$. }
\end{figure}

\begin{figure}[t]
	\subfigure[Ratio of Deviations]{\label{fig:s2_dev}
        \includegraphics[width=0.49\textwidth]{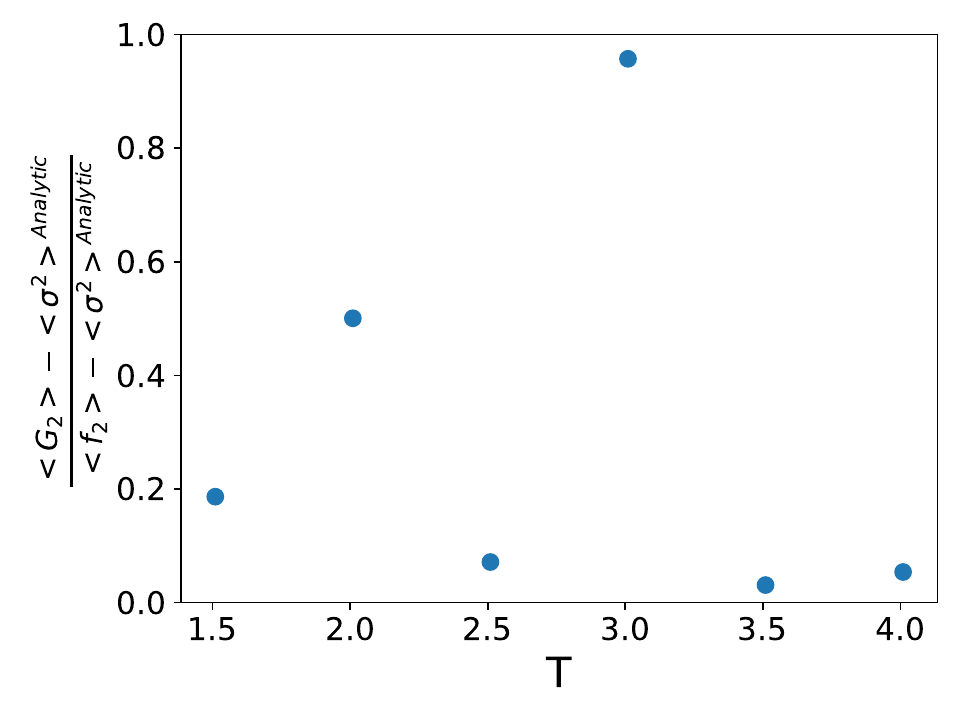}
			}
	\subfigure[Ratio of statistical errors]{\label{fig:s2_err}
				\includegraphics[width=0.49\textwidth]{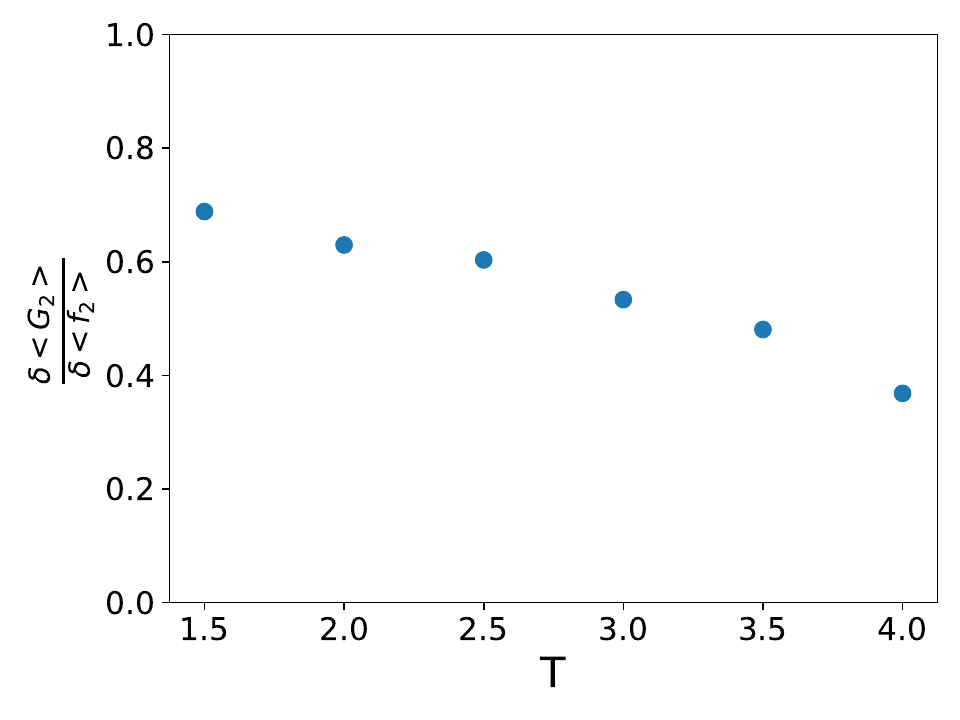}
	}
	\caption{Left: Ratio of the deviations from analytic solution on a 4x4 lattice for $h=0.2$. Right: Ratio of the statistical errors of $\expval{f_2}$ and $\expval{G_2}$ }
\end{figure}

The magnetization $\langle\sigma\rangle $ and the susceptibility $\chi$ can be written in terms of the total monomer number $M$ as follows,
\begin{align}
	\langle \sigma^{n} \rangle &= \frac{1}{(N\beta)^n}\frac{1}{Z} \frac{\partial^n Z}{\partial h^n} = \langle f_n \rangle \\
	\langle \sigma \rangle &= \tanh(\beta h) + \frac{\langle M \rangle}{\sinh(\beta h)\cosh(\beta h)} = \langle f_1 \rangle \\
	\chi &= \langle \sigma^2 \rangle - \langle \sigma \rangle^2 
=\frac{1}{N\cosh^2(\beta h)} -\frac{1}{N}\left(\frac{1}{\sinh^2(\beta h)}+\frac{1}{\cosh^2(\beta h)}\right)\langle M \rangle \nonumber\\
&\qquad \qquad \qquad \quad + \frac{1}{(\sinh(\beta h)\cosh(\beta h))^2} \left(\langle M^2 \rangle - \langle M \rangle^2\right) = \langle f_2 \rangle - \langle f_1 \rangle^2 \,.
\end{align}
Here, we define $f_n$ to distinguish the observables written in terms of $M$ from the same observable written in terms of the improved estimator $G_2$. For example, $f_2 = \sigma^2(M)$.
 Note that they are not the same $f_2 \neq G_2$ before ensemble averaging, as they have different distributions. 
We compare $\langle f_2 \rangle$ and $\langle G_2 \rangle$ with the exact solution in Fig.~\ref{fig:s2}. The analytic solution is subtracted in Fig.~\ref{fig:s2_sub} to see the statistical errors and the deviations from the analytic result. Since $\langle G_2 \rangle$ as a worm estimator has better statistics and hence smaller error bars compared to $\langle f_2 \rangle$ (see Fig~\ref{fig:s2_err}), also its mean value is closer to the analytic result, see Fig.~\ref{fig:s2_dev}.
This advantage of improved estimators concerning statistical errors is strong in particular at higher temperatures, including the vicinity of a critical point, but it weaker at low temperatures as the worm algorithm is less efficient here, since the average worm length is very large here.  

\section{Machine learning strategy}
\begin{figure}[hb]
	\subfigure[Training data]{
\label{fig:corr}
        \includegraphics[width=0.49\textwidth]{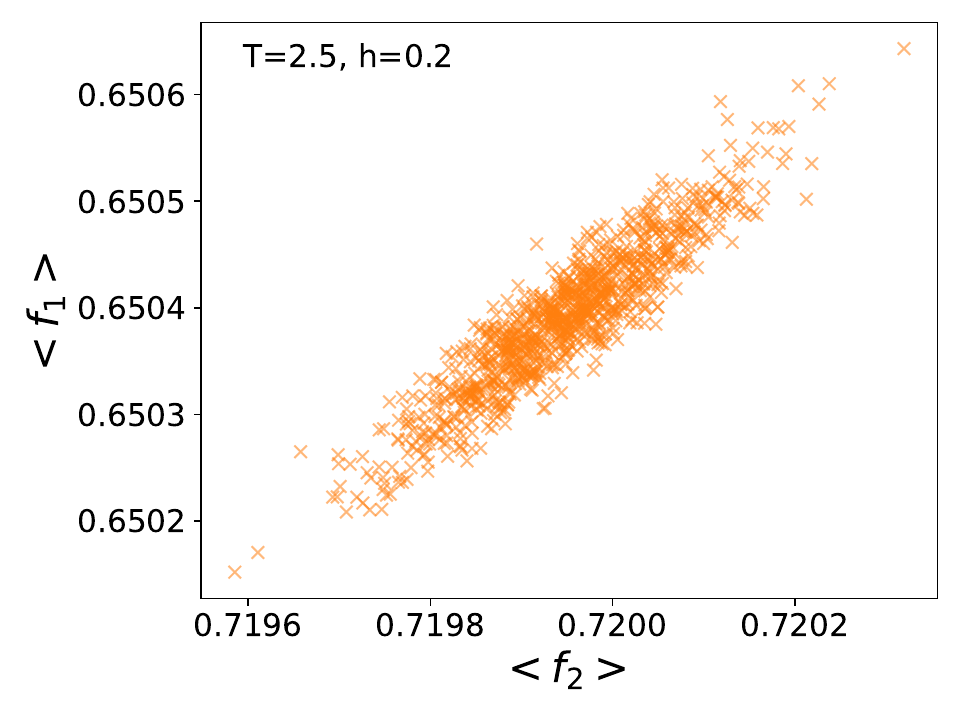}
			}
	\subfigure[Estimation by decision tree regression]{\label{fig:esti}
				\includegraphics[width=0.49\textwidth]{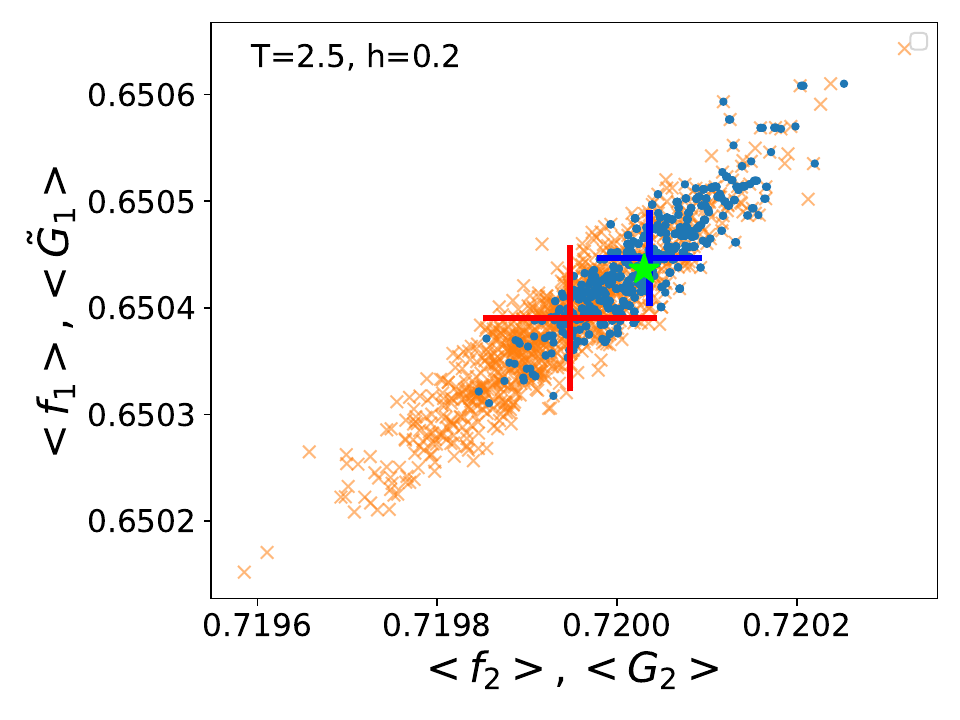}
	}
	\caption{Left: Correlation between the bootstrap samples of $\expval{f_2}$ and $\expval{f_1}$ at $T=2.5$ and $h=0.2$. Right: blue points are the decision tree regression prediction for $\expval{G_2}$ input. Green point is the analytic solution.}
\end{figure}

\begin{figure}[t]
	\subfigure[Histogram of $\expval{f_2}$ and $\expval{G_2}$]{
\label{fig:histo_s2}
        \includegraphics[width=0.49\textwidth]{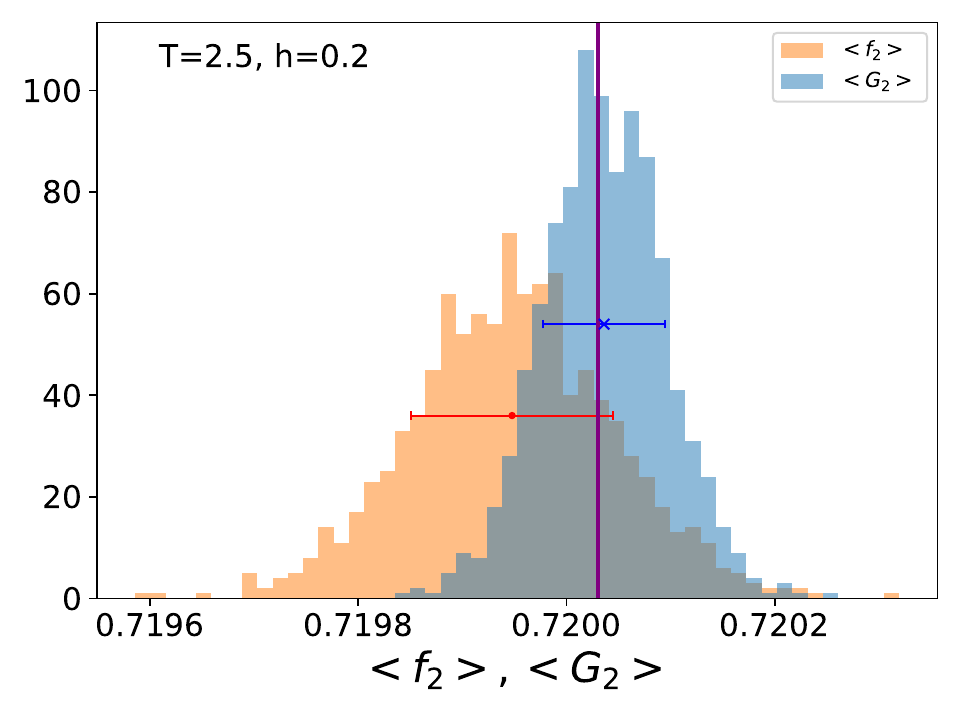}
			}
	\subfigure[Histogram of $\expval{f_1}$ and $\expval{\tilde{G}_1}$]{
	\label{fig:histo_s1}
				\includegraphics[width=0.49\textwidth]{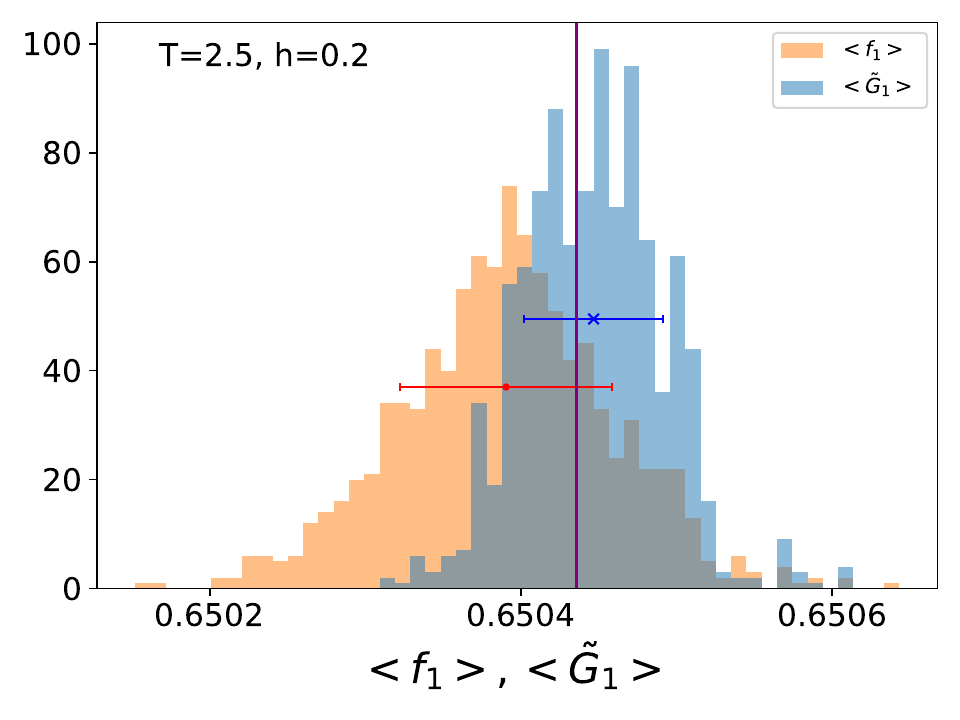}
	}
	\caption{Left: histogram of $\expval{f_2}$ and $\expval{G_2}$ at $T=2.5$ and $h=0.2$. Right: histogram of $\expval{f_1}$ and $\expval{\tilde{G}_1}$ at $T=2.5$ and $h=0.2$.}
\end{figure}

\begin{figure}[b]
	\subfigure[Magnetization $\expval{\sigma}$]{

        \includegraphics[width=0.49\textwidth]{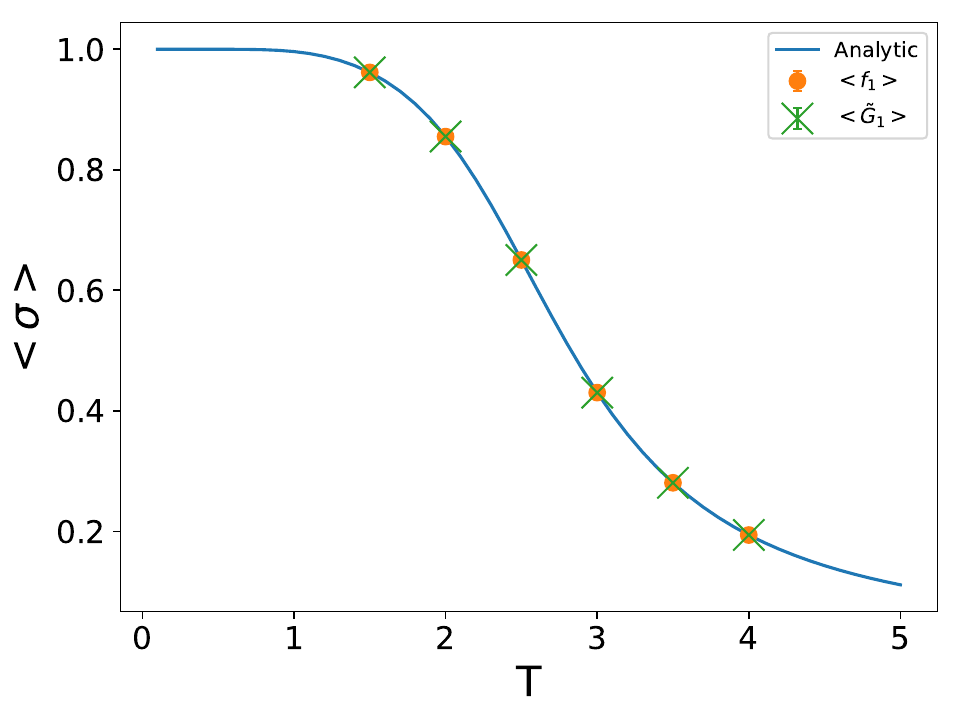}
			}
	\subfigure[Deviations]{
				\includegraphics[width=0.49\textwidth]{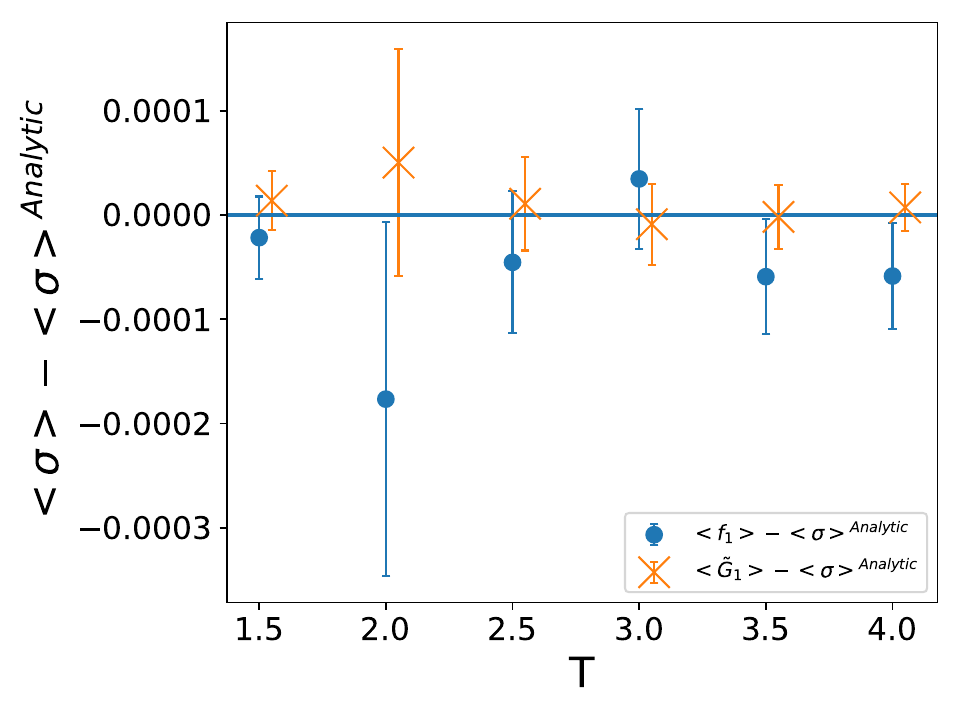}
	}
	\caption{\label{fig:s1} Left: Comparison of $\expval{f_1}$ and $\expval{\tilde{G}_1}$ with analytic solution on $4 \times 4$ lattice with external field $h=0.2$. Right: subtracting analytic solution from $\expval{f_1}$ and $\expval{\tilde{G}_1}$.}
\end{figure}

The important observable to determine the critical temperature is the susceptibility \linebreak $\chi=\expval{\sigma^2} -\expval{\sigma}^2$. The pseudo-critical temperature can be determined from its peak via finite size scaling. 
Whereas $\expval{\sigma^2}$ can be determined by $\expval{G_2}$, there is no improved estimator for $\expval{\sigma}$, and it has to be determined by $\expval{f_1}$, which is less accurate. 
While $\delta \expval{G_2}$ is small, the dominant error of the
susceptibility comes from $\delta (\expval{f_1}^2)$. 
The goal of our	 machine learning strategy is to reduce the statistical error of the susceptibility by predicting a new observable $\expval{\tilde{G}_1}$, which corresponds to $\expval{\sigma}$, with a reduced error. 
To obtain the general mapping of the distributions of the means $\expval{\sigma^2}$ and $\expval{\sigma}$, we consider as training data the correlation of bootstrap samples ($n=1000)$ between $\expval{f_2}$ and $\expval{f_1}$. With this, we train the
machine this correlation on a $4\times 4$ lattice with the external field $h=0.2$, as presented in Fig.~\ref{fig:corr}. 
In Fig.~\ref{fig:esti}, we present the machine learning prediction
$\expval{\tilde{G}_1}$. 
We obtain this mapping by applying the decision tree regression method from the scikit-learn library~\cite{scikit-learn}.
The decision tree regression method is used to select the closest data point of the input. 
The blue points are selected by the decision tree from the input of
$\expval{G_2}$.
The red and blue crosses indicate the statistical error of $\expval{f_2}$,
$\expval{f_1}$ and $\expval{G_2}$, $\expval{\tilde{G}_1}$.
Comparing with the analytic solution, the green star, the machine learning
prediction $\expval{\tilde{G}_1}$ has smaller statistical error and deviation
than $\expval{f_1}$.

The distribution of the bootstrap samples is Gaussian. 
In Fig.~\ref{fig:histo_s2}, we have shown that both $\expval{f_2}$ and
$\expval{G_2}$ are Gaussian distributions and $\expval{G_2}$ has smaller
statistical error. 
Here, the purple line is the analytic result. 
After applying the decision tree regression, the machine learning prediction
has also Gaussian distribution in Fig.~\ref{fig:histo_s1}. 

\begin{figure}[t]
	\subfigure[Ratio of deviations]{
\label{fig:s1_dev}
        \includegraphics[width=0.49\textwidth]{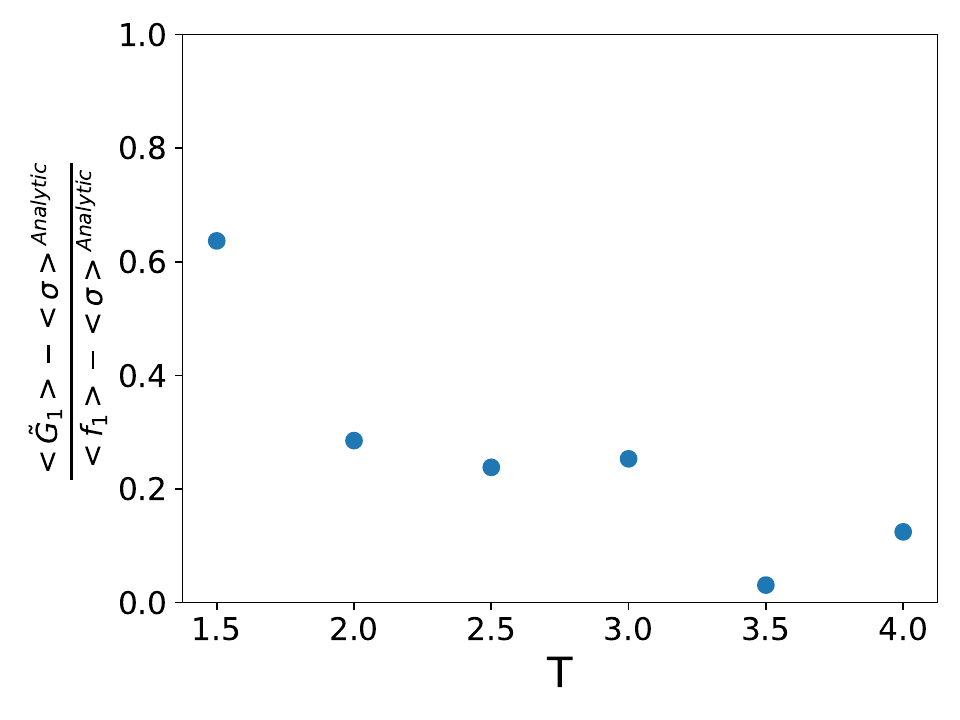}
			}
	\subfigure[Statistical error reduction]{\label{fig:s1_err}
				\includegraphics[width=0.49\textwidth]{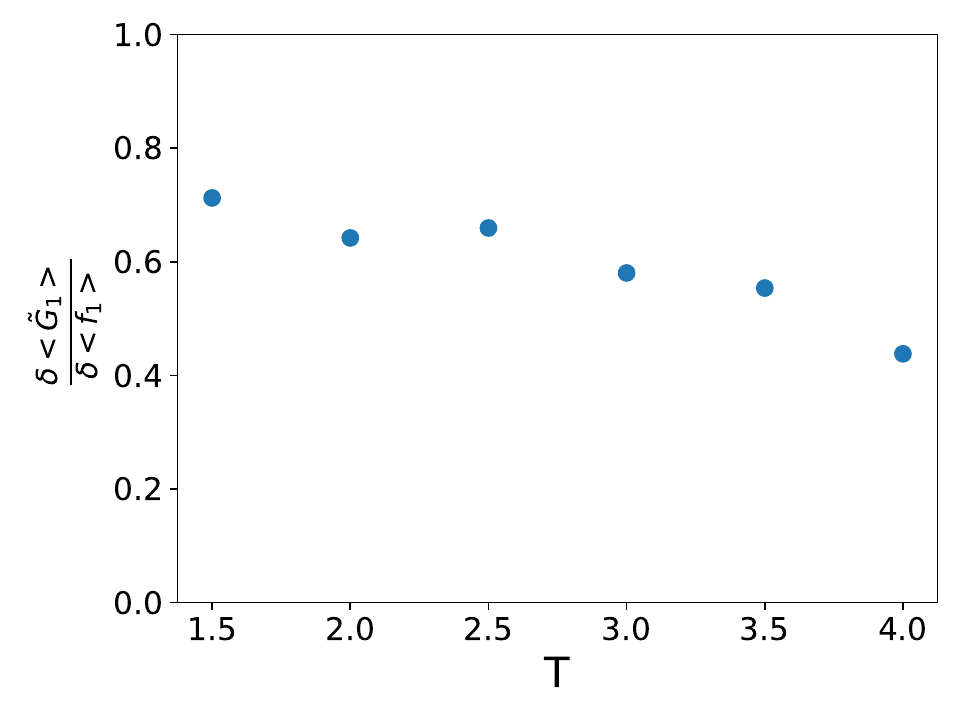}
	}
	\caption{Left: Comparison of $\expval{f_1}$ and $\expval{\tilde{G}_1}$ with analytic solution on $4 \times 4$ lattice. Right: Ratio of the statistical errors of  $\expval{f_1}$ and $\expval{\tilde{G}_1}$ }
\end{figure}

The magnetization $\expval{\tilde{G}_1}$ obtained by machine learning for
temperatures from $T=1.5$ to $T=4.0$ are presented in Fig.~\ref{fig:s1}. 
In Fig.~\ref{fig:s1_dev} and Fig.\ref{fig:s1_err}, we compare the deviation and statistical error of $\expval{\tilde{G}_1}$ and $\expval{f_1}$. As a result, machine learning predictions are more accurate and closer to the true result. The statistical errors are reduced by about 40\%. 

\begin{figure}[h]
	\subfigure[Susceptibility]{
        \includegraphics[width=0.49\textwidth]{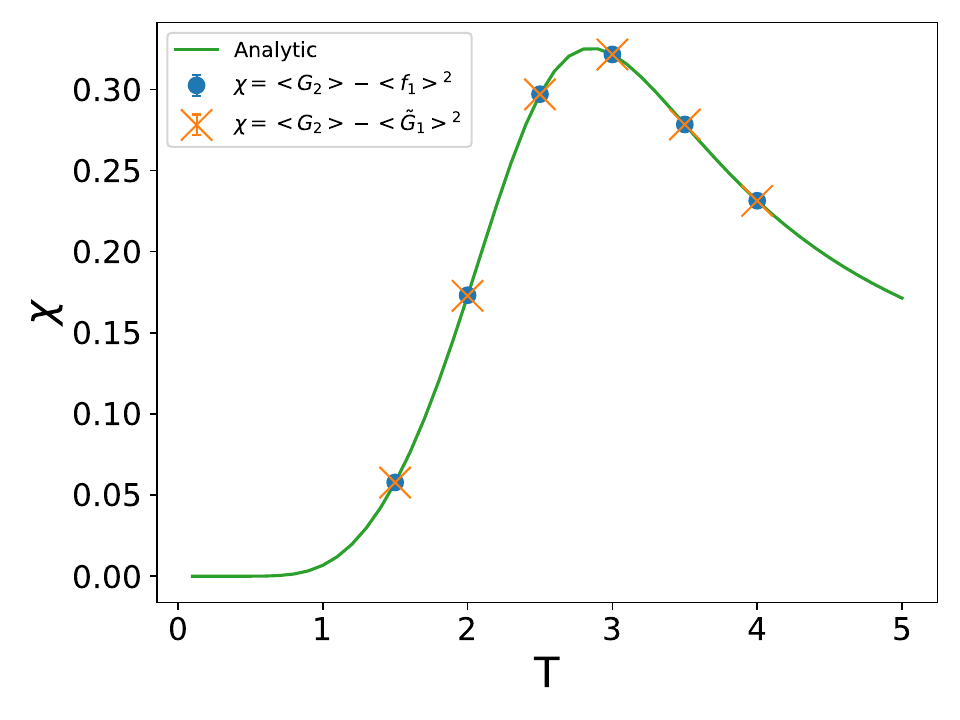}
			}
	\subfigure[Deviations]{
				\includegraphics[width=0.49\textwidth]{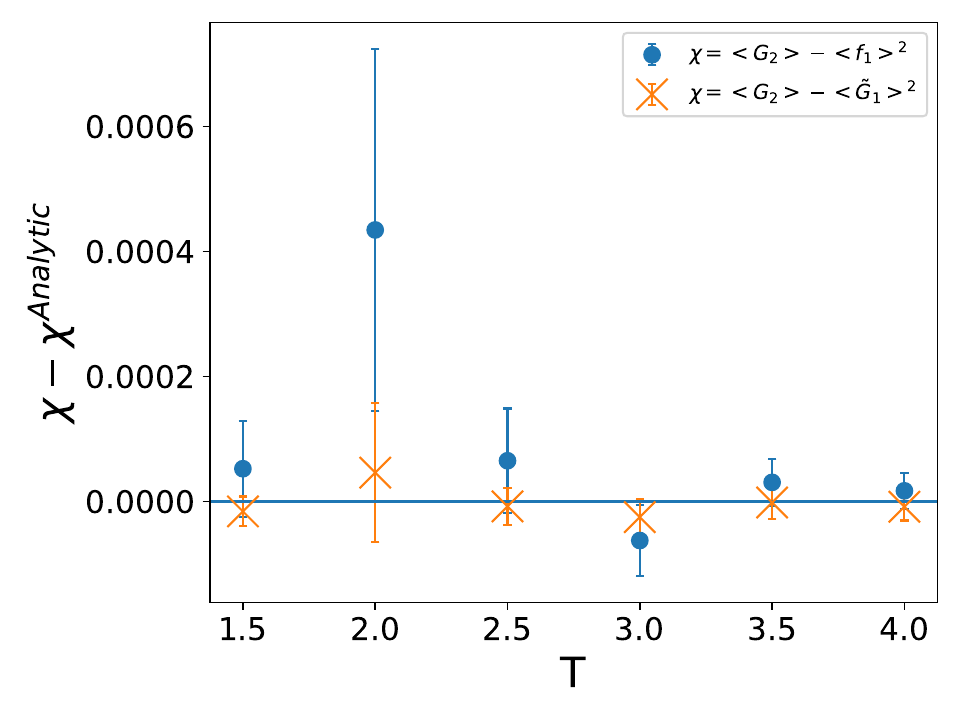}
	}
	\caption{
\label{fig:sus}
		Comparison of $\chi=\expval{G_2}-\expval{f_1}^2$ and $\chi=\expval{G_2}-\expval{\tilde{G}_1}^2$ with analytic solution on $4 \times 4$ lattice. }
\end{figure}

\begin{figure}[t]
	\subfigure[Ratio of deviations]{
\label{fig:sus_dev}
        \includegraphics[width=0.49\textwidth]{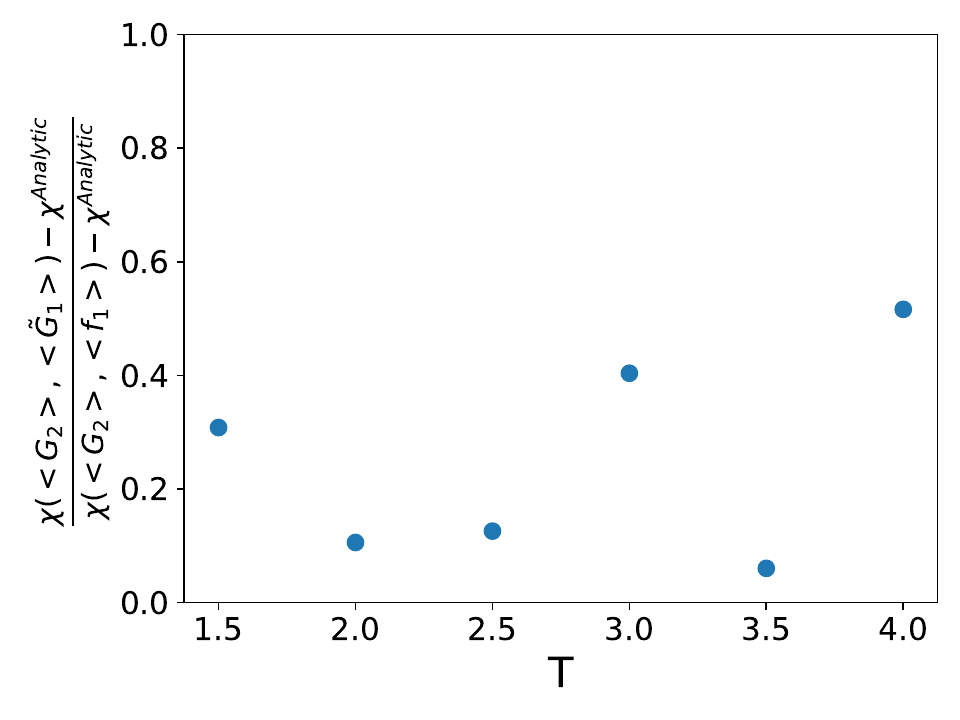}
			}
	\subfigure[Statistical error reduction]{\label{fig:sus_err}
				\includegraphics[width=0.49\textwidth]{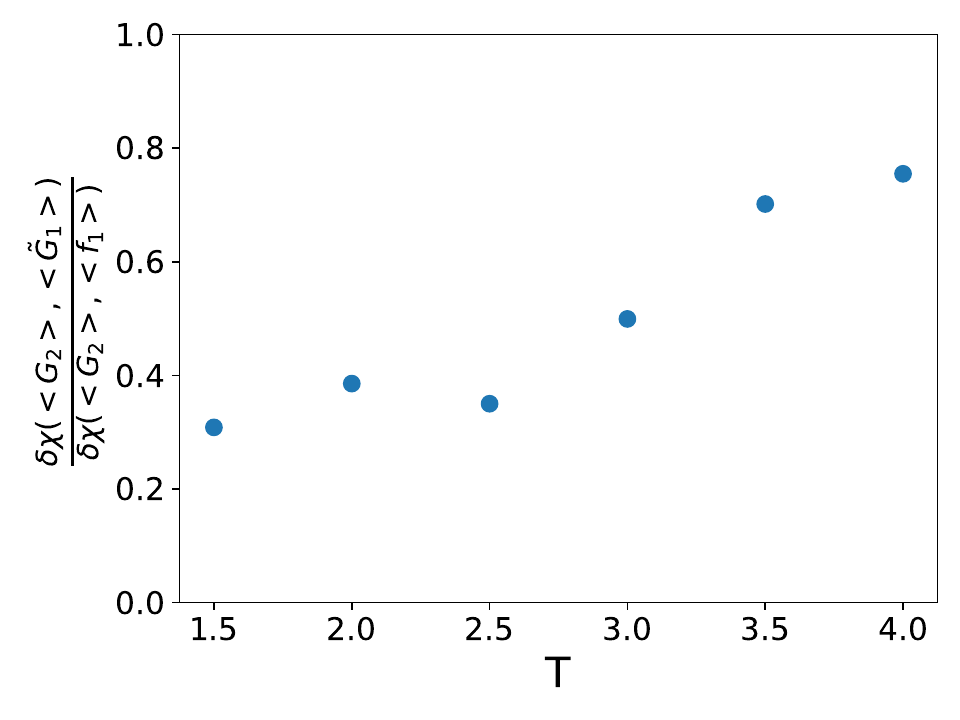}
	}
	\caption{Left: deviation from the analytic solution. Right: Statistical error reduction by machine learning method for susceptibility.}
\end{figure}

The susceptibility can be obtained by subtracting $\expval{f_1}^2$ or
$\expval{\tilde{G}_1}^2$ from $\expval{G_2}$. In Fig.~\ref{fig:sus}, we compare
two ways of calculations. 
When the machine learning prediction $\expval{\tilde{G}_1}^2$ is subtracted,
the results are closer to the analytic result. 
The reduction of deviations and statistical errors are presented in
Fig.~\ref{fig:sus_dev} and Fig.~\ref{fig:sus_err}. The statistical errors are
reduced by about 20-70\% depending on the temperature.
\section{Higher moments}
\begin{figure}[tb]
	\subfigure[Pearson Coefficients]{\label{fig:PearsonT}
				\includegraphics[width=0.49\textwidth]{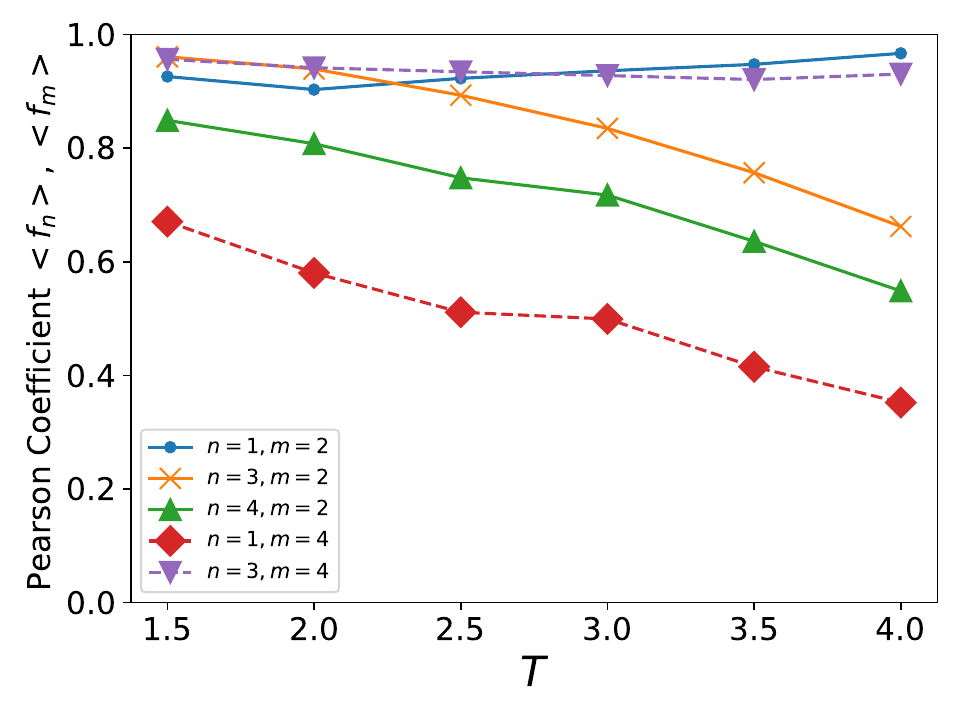}
	}
	\subfigure[Deviation from the analytic solution]{\label{fig:Pearson_stat}
	\includegraphics[width=0.49\textwidth]{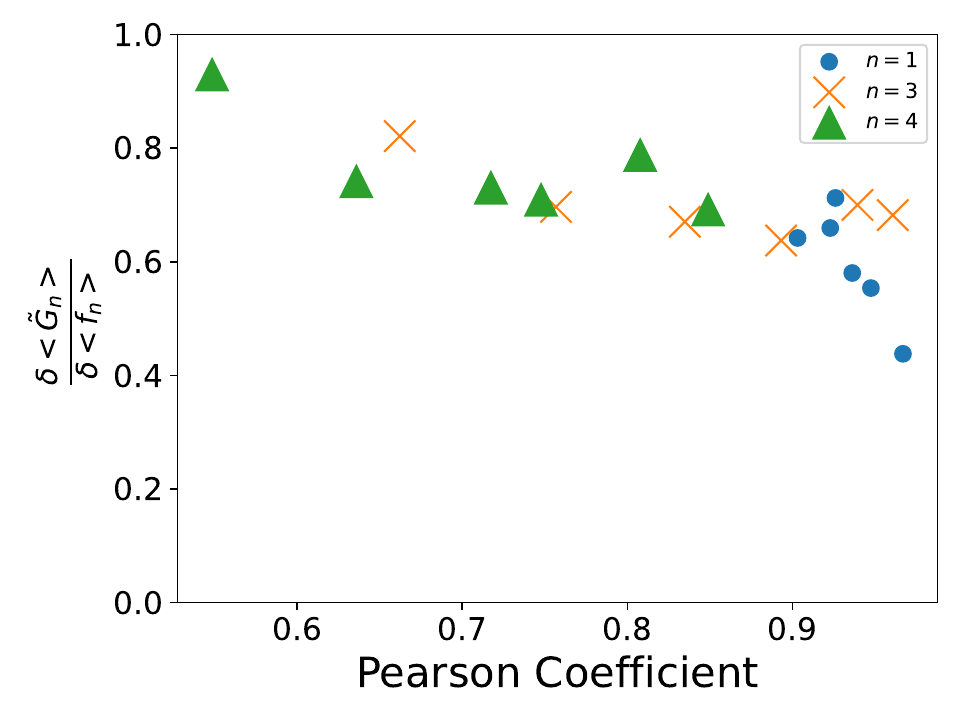}
			}
	\caption{
	Pearson coefficient of the correlation between $f_n$ and $f_m$, where $m=2,4$.
	Statistical error reduction with respect to Pearson coefficient.}
\end{figure}

This method is applicable to higher moment, for instance, $\expval{\sigma^3}$ or $\expval{\sigma^4}$.
The effect of the statistical error reduction depends on how strong the correlation is, which can be quantified by the Pearson coefficient. 
In Fig.~\ref{fig:Pearson_stat}, we show the dependence of statistical error reduction with respect to Pearson coefficient of the correlation between $\expval{f_2}$ and $\expval{f_n}$, where $n=1,3,4$.
In the case of $\expval{f_1}$, it is clear that the statistical error reduction is more effective at strong correlations. However, in the case of higher moments, $n=3,4$, the statistical errors are smaller but not as much as for the magnetization. 
The reason is that the accuracy of the input data affects the error reduction. In Fig.~\ref{fig:s2_err}, the input data is less accurate at lower temperatures. Hence the error reduction is less effective despite the Pearson coefficient is larger than $0.9$. The two data points in Fig.~\ref{fig:Pearson_stat} for $n=3$ with largest Pearson coefficient correspond to rather low temperatures $T=1.5, 2.0$. Their Pearson coefficients are close to one but the error reduction is not large, about 30\%. On the other hand, the data for $n=1$ with largest Pearson coefficient correspond to high temperature which has very accurate input data.  Despite of accurate input data at high temperatures,  the small error reduction for the other data is due to the weak correlation, as seen in Fig.~\ref{fig:PearsonT} for $n=3,4$ and $m=2$. 
The Pearson coefficient for the correlation between $\expval{f_3}$ and $\expval{f_4}$ is large even at high temperatures. Hence, if we have the accurate input $\expval{G_4}$, $\expval{\tilde{G}_3}$ can be evaluated by our machine learning method precisely. $\expval{G_4}$ can be sampled by introducing a second worm:
sampling the Ising model with such a two-worm algorithm~\cite{Brett:2015}, the four-point function $G(x,y,z,w)$ and its integrated expectation value $\langle G_4\rangle$ can be directly measured as an improved estimator. Then the correlation between $\expval{f_3}$ and $\expval{f_4}$ can be learned from the input $\expval{G_4}$, resulting in an error reduction for $\expval{\tilde{G}_3}$.

\section{Conclusion}
We developed an error reduction strategy using the decision tree method. We have tested this method for the Ising model in its dual representation and found that the error of the magnetization $\expval{\sigma}$ is reduced about 40\%. Moreover, for the susceptibility, the mean value of the machine learned prediction is closer to the analytic result.
Applying this method to higher moments is less efficient because of the weak correlation of the worm estimator with the higher moments in terms of monomers. A two-worm algorithm with external field is required for its improvement, and is currently under investigation.  We also plan to test this method for larger volumes and higher dimensions and determine observables for finite size scaling, such as the Binder cumulant, with reduced statistical error form the decision tree method.
Finally, we aim is to apply this method to strong coupling lattice QCD in the dual representation~\cite{Gagliardi:2017uag}. Here, we may benefit for improving on chiral and nuclear observables, to pinpoint the QCD phase diagram in the strong coupling regime.


\acknowledgments
J.K. was supported in part by the NSFC and the Deutsche Forschungsgemeinschaft (DFG, German Research Foundation) through the funds provided to the Sino-German Collaborative Research Center TRR110 "Symmetries and the Emergence of Structure in QCD" (NSFC Grant No. 12070131001, DFG Project-ID 196253076 - TRR 110)
W.U. acknowledges support by the Deutsche Forschungsgemeinschaft (DFG, German Research Foundation) through the CRC-TR 211 'Strong-interaction matter under extreme conditions'– project number 315477589 – TRR 211.
\bibliography{refs}
\end{document}